\documentclass[manuscript]{aastex}
\usepackage{subfig}

\shorttitle{The Origin of Gamma-Rays from Globular Clusters}
\shortauthors{Cheng et al.}

\begin{document}


\title{The Origin of Gamma-Rays from Globular Clusters}
\author{
K.S.Cheng\altaffilmark{1}, D. O. Chernyshov\altaffilmark{2}, V. A. Dogiel\altaffilmark{3}, C. Y. Hui\altaffilmark{4},  and A.K.H. Kong\altaffilmark{5} }

\altaffiltext{1}
{Department of Physics, University of Hong Kong, Pokfulam Road, Hong
Kong}
\altaffiltext{2}
{Moscow Institute of Physics and Technology, Institutskii lane, 141700
Moscow Region, Dolgoprudnii, Russia.}
\altaffiltext{3}
{I.E.Tamm Theoretical Physics Division of P.N.Lebedev
Institute, Leninskii pr, 53, 119991 Moscow, Russia}
\altaffiltext{4}
{Department of Astronomy and Space Science, Chungnam National University,
Daejeon, South Korea}
\altaffiltext{5}
{Institute of Astronomy and Department of Physics, National Tsing Hua University, Hsinchu, Taiwan}

\begin{abstract}
Fermi has detected gamma-ray emission from eight globular clusters.
We suggest that the gamma-ray emission from globular clusters may result from the
inverse Compton scattering between relativistic electrons/positrons in the pulsar
wind of MSPs in the globular clusters and background soft photons including cosmic microwave/relic photons, background star lights in the clusters, the galactic infrared photons and the galactic star lights. We show that the gamma-ray spectrum from 47 Tuc can be explained equally well by upward scattering of either the relic photons, the galactic infrared photons or the galactic star lights whereas the gamma-ray spectra from  other seven
globular clusters are best fitted by the upward scattering of either the galactic infrared photons or the galactic star lights. We also find that
the observed gamma-ray luminosity is correlated better with the combined factor of the encounter rate and the background soft photon energy density.
Therefore the inverse Compton scattering may also contribute to the observed gamma-ray emission from globular clusters detected by Fermi in addition to the standard curvature radiation process.
Furthermore, we find that the emission region of high energy photons from globular cluster produced by inverse Compton scattering  is substantially larger than the core of globular cluster with a radius $>$10pc. The diffuse radio and X-rays emitted from globular clusters can also be produced by synchrotron radiation and inverse Compton scattering respectively. We suggest that future observations including radio, X-rays, and gamma-rays with energy higher than 10 GeV and better angular resolution can provide better constraints for the models.

\end{abstract}



\section{Introduction}
Globular clusters (GCs) are the most dense stellar system,which results in frequent dynamical interactions.
In particular the formation rate per unit mass of Low Mass X-ray Binaries (LMXBs) is orders of magnitude higher
in GCs than in the Galactic field (Katz 1975; Clark 1975). It is generally believed that LMXBs are progenitors of
millisecond pulsars (MSPs) (e.g. Alpar et al. 1982). Therefore it is not surprised that 80\% of detected
MSPs are located in GCs. So far 140 MSPs have been detected in 26 GCs.\footnote{http://www2.naic.edu/$\sim$pfreire/GCpsr.html}

With the launch of the Fermi Gamma-ray Space Telescope, we have entered a new
era of high energy astrophysics. As the sensitivity of the Large Area Telescope
(LAT) on the spacecraft is much higher than that of EGRET, it has already
led to many interesting discoveries including the detection of GeV gamma-rays from GCs.
Shortly after the detections of two GCs, i.e. 47 Tuc (Abdo et al. 2009) and Terzan 5 (Kong et al. 2010), with GeV
gamma-ray emission, other six GCs have
been identified as gamma-ray emitters(cf. Abdo et al. 2010a; Abdo et al. 2010b).

It is generally believed that the gamma-ray emission from GCs either comes from magnetospheres
of MSPs or produced by inverse Compton scattering between electrons accelerated in the relativistic pulsar wind
and background soft photons. In fact before the detection of gamma-rays from GCs, Wang et al. (2005) have
shown that the
curvature radiation spectrum calculated from the outergap model of Zhang \& Cheng (1997) produced from unresolved MSPs
in the galactic center can result in a simple power law with an exponential cut-off energy at $\sim 3$~GeV.
They used the observed
distribution functions of MSPs from the field, from the GCs and the combination of these two distributions, and they found that the model spectrum was quite consistent with the
diffuse gamma-ray spectrum detected by EGRET in the direction of the galactic center. However it is important to note that the total gamma-ray
spectra calculated from these three different distributions (cf. Fig. 4 of Wang et al. 2005) are actually very similar. Therefore it is
very difficult to constraint the models by using average spectrum.
Recently Venter \& de Jager (2008) and Venter et al. (2009) calculated the expected
flux of gamma-rays produced by the curvature radiation of
electrons in pulsar magnetospheres. Venter \& de Jager (2008) first calculated the expected GeV flux from 47 Tucanae (47 Tuc) by using an unscreened (pair-starved polar cap) electric field (see e.g. Harding, Usov, \& Muslimov 2005) for 12 out of the 13 MSPs they considered, and the screened field for only 1 MSP with a relatively high spindown power based on the approximation of
a screened electric field by Dyks \& Rudak
(2000).  Venter et al. (2009) extended the model to include the inverse Compton component, which can produce TeV photons. Their model predictions are consistent with the later reported results by Fermi (Abdo et al. 2009) but the predicted TeV flux seems to be higher than the observed upper limits for 47 Tuc (Aharonian et al. 2009). On the other hand the total number of millisecond pulsar is unclear and it is still possible that by adjusting the model parameters both GeV and TeV observed results can be explained in this model.

However, the radio and X-ray properties of MSPs in GCs are found to be rather different
from those located in the Galactic field (Bogdanov et al. 2006; Hui, Cheng \&
Taam 2009, 2010). The difference can be possibly related to the complicated multipole magnetic
field structure of the MSPs in a cluster, which is a consequence of frequent
stellar interaction (cf. Cheng \& Taam 2003 and see \S2 for a more detailed account).
In fact the complicated surface magnetic field structure
can have a very dramatic effect on both polar gap and outer gap structure. If the surface local magnetic field of
millisecond pulsars is of order of $10^{12}$G as suggested by Ruderman (1991), Cheng \& Zhang (1999) showed that the polar
gap potential drop can reduce to $10^{11}$V, which makes GeV-photon production become very difficult
whereas large number of pairs can still be produced via magnetic pair creation process.
Another consequence of
complicated surface magnetic field is to turn off the outergap. Ruderman \& Cheng (1988) argue that if the open field lines
are curving upward due to the effect of local field then in this case electron/positron pair production
and outflow can occur on all open field lines. Consequently the outer magnetospheric gap is quenched by these pairs.
Furthermore Cheng and Taam (2003) have also pointed out that most X-ray spectra of pulsars in 47 Tuc can be described by a thermal spectrum with a characteristic temperature insensitive to the pulsar parameters resulting from the fact that the surface magnetic field structure of MSPs in globular clusters should be dominated by complicated multiple field structure and consequently the polar gap is substantially suppressed and outergap should not exist. We also explain why the X-ray luminosity of MSPs in the globular clusters and MSPs in the field obey different relation with spin-down power. Zalvin (2006) and Bodanov et al. (2006) both conclude that the spectral properties of the MSPs in the field and in GCs are found to be different. We have found some good reasons to believe that properties of MSPs in globular clusters differ from MSPs in the field. It should be noticed that the spectra of almost all the Fermi-LAT pulsars including MSPs, except very young pulsars like the Crab pulsar, can be explained in terms of CR mechanism. Other models even they can fit the Fermi data of globular clusters equally well but they cannot be accepted as alternative models unless they have other new predictions and are supported by observations.
Nevertheless with all these observational hints, we propose that there may be an alternative/additional emission mechanism to produce the observed gamma-rays detected by Fermi-LAT and explore the new predictions from this model.

Bednarek \& Sitarek (2007) analyzed gamma-ray emission
of electrons accelerated  at  shock waves originated in collisions
of the pulsar winds and/or  inside the pulsar magnetospheres when
gamma-rays are generated by the inverse Compton (IC) scattering of
ultra-relativistic electrons of relic and stellar photons. Both of these models can give reasonable
explanation for the gamma-ray emission from 47 Tuc. It should be noticed that both of these models predict that
gamma-rays are emitted from core region of GC, i.e. $<1$ pc, where most MSPs are located.
The key difference between these two classes of models is that the inverse Compton model
predicts the existence of very high energy gamma-rays, which can be detected by MAGIC and HESS.

In this paper we also study the inverse Compton scattering between the relativistic electrons/positron in the pulsar wind and the background soft photons. We adopt the pulsar wind model proposed by Cheng, Taam \& Wang (2004, 2006).
To generalize the soft photon field in our investigation, in addition to relic photons
and the star light photons in the GC, we include the background soft photons from the galactic disk including the infrared photons and star light photons of the galactic disk. Our calculations do not restrict the inverse Compton scattering only in the core instead we extend our calculation to several hundred pc from the core of GC. By the fitting the observed data of 47 Tuc and Terzan 5, our conclusion is significantly different from previous findings. The paper is organized as follows. In section 2, we summarize the observations of 47 Tuc and Terzan 5. In section 3, we describe the pulsar wind model by Cheng, Taam \& Wang (2004, 2006). In section 4 we present the spatial dependent inverse Compton scattering model. In section 5 we apply our model to explain the data of 47 Tuc, Terzan 5 and other six globular clusters observed by Fermi. In section 6 we discuss how other energy bands can constrain various IC models. We summarize our model predictions including a simple correlation analysis between gamma-ray luminosity and the background soft photon density in section 7.

\section{Observational properties of $\gamma-$ray emitting GCs}\label{s_obs}
\subsection{47 Tuc}
Apart from the high collision frequency that due to the high stellar density inside the cluster,
the relatively high metal content in 47~Tuc
can further facilitate the formation of binaries with more
efficient magnetic braking  (cf. Ivanova 2006).  Therefore,
a large binary population is expected in 47 Tuc.

With a deep X-ray survey of Chandra Observatory, 300 X-ray
sources within its half-mass radius have been revealed from 47 Tuc
(Heinke et al. 2005). This population contains various classes of
exotic binaries, including cataclysmic variables (CVs),
chromospherically active binaries (ABs), LMXBs
as well as MSPs. On the other hand, dedicated
radio survey have so far uncovered 23 MSPs in this cluster
(Camilo et al. 2000) which have reached a detection threshold of
$\sim0.5$~mJy~kpc$^{2}$. Among these 23 MSPs, 19 of them have
their X-ray counterparts been identified (Bogdanov et al. 2006).

The X-ray luminosities of the 47~Tuc pulsars are
in the range of $L_{X}\sim10^{30}-10^{31}$ ergs~s$^{-1}$
(Bogdanov et al. 2006). The X-ray spectra of the majority of these
pulsars can be well-described by a thermal model (blackbody or
or neutron star hydrogen atmosphere model) with the temperature
$T_{\rm eff}\sim(1-3)\times10^{6}$~K and the emission radius
$R_{\rm eff}\sim0.1-3$ km (Bogdanov et al. 2006).
These properties are found to be very different from the MSPs
in the Galactic field (cf. Hui et al. 2009 and references therein).
While the MSPs in 47 Tuc are essentially thermal emitters (except
for some have intrabinary shock observed such as 47 Tuc W), the
MSPs in the Galactic field generally require two components to model
their X-ray spectra which includes a hot polar cap component plus a
non-thermal power-law tail (Zavlin 2006).

To account for the differences between the 47 Tuc MSPs (or generally
the MSPs in GCs) and the MSP population in the Galactic field, it has been suggested that
the absence of non-thermal X-ray from
the cluster MSPs can possibly related to the complicated multipole magnetic
field structure(cf. Grindlay et al. 2002; Cheng \& Taam 2003). Because of frequent stellar interaction, MSPs in a
GC can possibly change their companion several times throughout their
lives. As the orientation of the binary after each exchange can differ,
the direction of the angular momentum accreted during the mass transfer
phase subsequent to each exchange can vary possibly affecting the magnetic
field configuration at the neutron star surface.  Such an evolution could
lead to a much more complicated multipole magnetic field structure for the MSPs in
the GCs than in the case in the Galactic field. In such a complicated magnetic
field Ruderman \& Cheng (1988) have argued that high energy curvature
photons will be emitted and subsequently converted into pairs to quench
the accelerating region. This provides an explanation for the absence of
non-thermal emission in the cluster MSPs. For the same reason,
the complicated multipole magnetic field structure can also possibly alter the coherent
radio emission and provide the explanation for the different radio luminosity
distribution of the cluster
MSPs in comparison with that of the disk MSP population (Hui, Cheng, \& Taam 2010).

Apart from the magnetospheric emission, it has long been speculated that
pulsar wind nebulae (PWN) from the MSPs can have possible contribution in a GC.
In 47 Tuc, the low dispersion measure for its MSP population (Freire et al. 2001)
suggests that some mechanism operates to reduce the mass of gas in the central region
expected to be accumulated in the $\sim10^{7}-10^{8}$ years interval between passages
of the cluster through the Galactic disk (cf. Camilo \& Rasio 2005). The
outflow accompanying the relativistic winds from the MSPs in the cluster
could possibly reduce the amount of intracluster gas (cf. Spergel 1991).
Motivated by this insight, Hui, Cheng \& Taam (2009) have systematically
searched for the X-ray signature of pulsar wind nebulae within the cores of
a group of GCs. However, there is no compelling evidence for any nebular
emission can be found in the cluster cores. In contrast, some MSPs in the field have already been
found to associate with PWNs, e.g. Hui \& Becker (2006); Stappers et al. (2003).
This non-detection has further
suggested that the emission properties of the MSP population in GCs are
intrinsically different from those of the MSP in the Galactic field.

For further investigating the differences between these two populations,
gamma-ray observations can provide us important information for this study.
Shortly after the operating of LAT, the gamma-ray emission ($>200$~MeV) from
47~Tuc has been detected and this is the first time that a GC is detected in this
high energy regime (Abdo et al. 2009). The gamma-ray photons from 47 Tuc are
presumably the collective contribution by its MSP population. Its gamma-ray
spectrum can be well fitted by an exponentially cut-off power-law model with
a photon index of $\Gamma=1.3\pm0.3$ and a cut-off energy of
$E_{c}=2.5^{+1.6}_{-0.8}$~GeV (Abdo et al. 2009). The energy flux in $0.1-10$~GeV
is found to be $2.5\times10^{-11}$~ergs~cm$^{-2}$~s$^{-1}$ (Abdo et al. 2009). For
a distance of $\sim4$~kpc, the gamma-ray luminosity has put an upper bound for
the MSP population in 47 Tuc of 60 (Abdo et al. 2009).

\subsection{Terzan 5}

Terzan 5 holds the largest MSP population among all the MSP-hosting
GCs. Currently, there are 33 pulsars that have been found in Terzan 5
(see Ransom et al. 2005; Hessels et al. 2006). It has been shown that two body encounter rate plays
an important role in the formation of low-mass X-ray binaries in globular clusters (Verbunt \& Hut 1987;
Verbunt et al. 1989). Since both collision
frequency and the metallicity of Terzan~5 are even higher than the
values found in 47~Tuc, a larger binary content
is expected in Terzan~5. Furthermore Fruchter \& Goss (1990, 2000) have identified strong
diffuse radio emission from Terzan 5. By using standard pulsar luminosity function, they estimate
that there are 50 MSPs, which beam toward Earth and imply 500-2000 MSPs in this GC.
By using the cumulative radio luminosity distribution function
Hui et al. (2010) have recently
predicted that the MSP population in Terzan~5 can be $\sim4-5$ times
higher than that in 47~Tuc. Because of the large number of MSPs,
it is expected to have strong $\gamma$-ray emission.
With data obtained in a $\sim17$ months of continuous observation by
LAT, the expected $\gamma$-ray emission from Terzan~5 have been
eventually detected at a significance level of $\sim27\sigma$
(Kong et al. 2010). The energy spectrum of Terzan 5 is best
described by an exponential cutoff power-law model, with a photon
index of $1.9\pm0.2$ and a cutoff energy at $3.8\pm1.2$ GeV.
The energy flux in $0.5-20$~GeV is found to be
$(6.8\pm2.0)\times 10^{-11}$ ergs cm$^{-2}$ s$^{-1}$. For comparison
with result reported for 47~Tuc, the flux in $0.1-10$~GeV is
$\sim1.2\times10^{-10}$~ergs cm$^{-2}$ s$^{-1}$.

The large
reservoir of MSPs in Terzan 5 could also provide the seed electrons
for inverse Compton scattering of star-light photons or non-thermal
bremsstrahlung emission from the deflection of the electrons by
interstellar medium. Recently, {\it Chandra} observation of Terzan 5
reveals extended diffuse X-ray emission outside the half-mass radius
of the cluster. The diffuse emission can be described by a steep
power-law with a photon index of 0.9 (1--7 keV) and it is likely to be
non-thermal in origin (Eger et al. 2010).


Comparing the gamma-ray properties of 47 Tuc with those of a recently discovered
gamma-ray emitting GC Terzan 5 (Kong et al. 2010), we found that there are certain
dissimilarities between these two GCs. First, despite the fact that it has been
suggested that Terzan 5 locates at a further distance than 47 Tuc, the
gamma-ray flux observed from Terzan 5 is $\sim5$ times higher than that of 47 Tuc.
Assuming the distance to Terzan 5 is at $\sim$6 kpc (Kong et al. 2010), instead of 10kpc, and the distance to 47 Tuc is 4 kpc, which implies
the gamma-ray luminosity of Terzan 5 is $\sim$12 times of 47 Tuc. If the properties of MSPs in these two clusters are similar
and the radiation mechanism is CR,
it implies the number of MSPs in Terzan 5 is 12 times of that of 47 Tuc. The observed ratio of millisecond pulsars is only $\sim$1.5,
so this required ratio seems to be unlikely. On the other hand if the radiation mechanism is
inverse Compton (IC) of the background soft photons, then the energy density of background photons is another factor to affect the gamma-ray luminosity. According to Strong and Moskalenko (1988), the soft photon densities
in Terzan 5 is a factor of $\sim$7 in optical and a factor of $\sim$5 in Infrared higher than that of that of 47 Tuc. Instead of a factor 12, the IC model only requires Terzan 5 has $\sim$5 times more MSPs than that of 47 Tuc, which is consistent with the prediction by Hui et al. (2010).
Second, the gamma-ray spectrum of 47 Tuc is found to be flatter than that of
Terzan 5, $\Gamma=1.9\pm0.2$ (Kong et al. 2010). Third,
there is an indication of an excess of $\gamma-$rays with
energies $>10$~GeV in Terzan 5 with a detection significance of $3.7\sigma$
(see Figure 1 in Kong et al. 2010).  In the case of 47 Tuc there is no
hint of any excess. These spectral differences may not be easily explained in terms of a simple CR radiation process whereas
the IC model, which also depends on an external factor, i.e. the background soft photon energy density, is more flexible to explain various spectral features.

\section{Pulsar Wind Model}
Rees \& Gunn (1974) prosposed a theoretical description of interaction between pulsar and its nebula.
They suggested that the central pulsar can generate a highly relativistic particle dominated wind that passes
through the medium in the supernova remnant, forming a shock front. The electrons and positrons in the shock are
envisioned to be accelerated to a power-law energy distribution and to radiation synchrotron radiation in the downstream region.
However, it is unlikely that electrons/poistrons can carry away all the spin-down power of pulsars near the light cylinder.
Kennel \& Coroniti (1984) have
introduced a magnetization parameter, $\sigma = \frac{B^2}{4\pi n\gamma_w mc^2}$, where $B$ is the magnetic field, $n$ is the particle number density, $\gamma_w$ is the Lorentz factor of relativistic particles in the wind and $m$ is the particle mass. In order to explain the observed radiation properties in the Crab nebula, $\sigma \sim 0.003$.  $e^{\pm}$ pairs are produced inside the
light cylinder in the polar gap (e.g. Ruderman \& Sutherland 1975; Fawley, Arons \& Scharlemann 1977) and/or outergap (e.g. Cheng, Ho \& Ruderman 1986). When electrons/positrons leave the light cylinder, they can only carry a very small fraction of spin-down power, which implies $\sigma>>1$. Therefore the magnetization parameter of pulsar wind must evolve from high-$\sigma$ to low-$\sigma$ in the down-stream. Coroniti (1990) has shown that the pulsar spin-down power initially carried away by low frequency electromagnetic waves can be converted into particle kinetic energy via magnetic reconnection process before reaching the shock radius.

Cheng, Taam \& Wang (2004, 2006) studied the nonpulsed X-ray emission of rotation-powered pulsars, they found that the nonpulsed X-ray luminosity
($L_x^{npul}$) is proportional to the pulsar spin-down power ($L_{sd}$) as $L_x^{npul} \propto L_{sd}^{1.4\pm 0.1}$. They argued that the nonpulsed X-rays should be emitted by the pulsar wind in the shock radius via synchrotron radiation. They used the simple one-zone model developed by Chevalier (2000) to estimate the relation between the spin-down power and nonpulsed X-ray luminosity. They assumed that if most spin-down power is eventually converted into the kinetic energy of protons and the proton current equals the Goldreich-Julian current$\dot{N}_{GJ}$ (Goldreich \& Julian 1969), then the Lorentz factor $\gamma_w$ of the
pulsar wind before it reaches the shock region can be expressed as
\begin{equation}
\gamma_w = 2\times 10^5 L_{34}^{1/2},
\end{equation}
where $L_{34}$ is the spin-down power in units of $10^{34}erg/s$(Cheng, Taam \& Wang 2004, 2006). With this simple estimation they obtained $L_x^{npul} \propto L_{sd}^{p/2}$, where $p$ is the power-law index of electron/positron in the shock region. In general the pulsar wind should consist of protons and  $e^{\pm}$ pairs. Assuming that the spin-down power of pulsar is still carried away by particle kinetic energy, i.e.
\begin{equation}
L_{sd} = \gamma_w \dot{N}_{GJ} m_p c^2 f_{e^{\pm}},
\end{equation}
where we assume that positive charges and negative charges are moving with same speed, $m_p$ is the proton mass, $f_{e^{\pm}}= 1 + \frac{m_e \eta_{e^{\pm}}}{m_p}$ and $\eta_{e^{\pm}}=\frac{\dot{N}_{e^{\pm}}}{\dot{N}_p}$ is number ratio between $e^{\pm}$ pairs and protons. By taking $e^{\pm}$ pairs into account, the Lorentz factor of pulsar wind becomes
\begin{equation}\label{ginj}
\gamma_w = 2\times 10^5 f_{e^{\pm}}^{-1} L_{34}^{1/2}.
\end{equation}
The value of  $\eta_{e^{\pm}}$ is model dependent. In the polar gap model,  $\eta_{e^{\pm}} \sim 10^2$ (e.g. Ruderman \& Sutherland 1975) which gives $f_{e^{\pm}} \sim 1$. On the other hand, in the outergap model this ratio $\eta_{e^{\pm}}$ is easily larger than $m_p/m_e$ (e.g. Cheng, Ho \& Ruderman 1986; Cheng \& Zhang 1999). However, the exact value of this ratio also depends on details of different outergap models. For example Wang et al. (2006) based on the MSP outergap model proposed by Zhang \& Cheng (2003) estimate the rate of electron/positron pairs produced by a MSP with $B=3\times 10^8$G and $P=3ms$ to be approximatelly equal to $5\times 10^{37}e^{\pm}s^{-1}$, which gives  $f_{e^{\pm}}\sim 30$. However, they have assumed all pairs produced near the neutron surface, which are moving inward initially, can be reflected by magnetic mirroring effect and escape through the open field lines, therefore their estimate should be an upper limit. Although the exact value of $f_{e^{\pm}}$ depends on the model details, in general it should be roughly $\sim 1-10$. The fraction of spin-down power carried away by $e^{\pm}$ pairs is given by
\begin{equation}
L_{e^{\pm}} = \frac{f_{e^{\pm}}-1}{f_{e^{\pm}}}L_{sd}=\zeta_{e^{\pm}}L_{sd}.
\end{equation}
Therefore the efficiency of spin-down power $\zeta_{e^{\pm}}$ carried away by pairs is roughly between 0.1 to 1. As suggested by Cheng \& Taam (2003), outergap may not exist for some MSPs with complicated surface magnetic field in globular clusters, if this is true then $\zeta_{e^{\pm}} \sim 0.1$ for those MSPs without outergap.
We would like to emphasize again that even outer gap does not exist large number of pairs can still be produced by the polar gap. In Ruderman \& Sutherland model, the pair multiplicity is typically $\sim 10^2$. Therefore in general number of pairs is still much higher than number of protons.

It is interesting to ask how much pulsar wind energy will lost in the shock region. In case of the Crab nebula, most of spin-down of pulsar are radiated within the nebula region. However the unpulsed X-ray luminosity of pulsars, which is assumed to be emitted from the shock region, is only a small fraction of the spin-down power. Furthermore Hui, Cheng \& Taam (2009) have tried to identify the diffuse X-rays of globular clusters resulting from pulsar wind shock regions, they conclude that there is no evidence that the diffuse X-rays can result from pulsar wind shock regions. They argue that the formation of shock region in globular cluster may be very difficult because the characteristic shock radius is much larger than the characteristic separation of stars due to the very low number density in the globular clusters.
In this paper we shall assume that pairs can be accelerated by absorbing the low frequency electromagnetic wave energy produced by the dipole radiation of pulsars to relativistic speed. If indeed the shock does not exist or very weak, pairs emitted and accelerated by pulsars can be treated as mono-energetic particles with a Lorentz factor given by Eq. (3). Since the particle energy loss in the shock is negligible, the pulsar spin-down power carried away by the pairs is given by Eq. (4). However, it is important to note that when pairs diffuse away from the globular clusters, they suffer inverse Compton energy loss by scattering with the background soft photons. Hence a simple power law with the energy
index $\Gamma_e \sim 2$ would be developed at a distance when the diffuse time equals the cooling time (Blumenthal \& Gould 1970).

\section{Inverse Compton Model}

One of the main differences between the CR and IC models  is the size of emitting region. The
curvature radiation of pulsars  is emitted from the central region of the GC
whose radius is about several pc. Such a tiny region cannot be
resolved by gamma-ray telescopes and, therefore, it is observed in
the gamma-ray range as a point-like source. On the other hand,
electrons/positrons ejected by pulsars may fill an extended region around
GCs and their IC radiation  is observed in this case as an
extended source. In this case  we should calculate the
spectrum and spatial distribution of electrons/positrons around the GC in
order to estimate the IC component of gamma-rays.

As usual, cosmic ray propagation in the interstellar medium is described
as diffusion process (see for details e.g. Berezinskii et al. 1990). The equations
for the distribution function of electrons $f({\bf r},E)$ has a
standard form
\begin{equation}\label{pr_state}
  \frac{\partial f}{\partial t}  - \nabla \left(D(r)\nabla f \right)+
  \frac{\partial}{\partial E}\left( \frac{dE}{dt} f\right) =
  Q(E,{\bf r},t)\,,
\end{equation}
where $dE/dt\equiv b(E, r)$ is the rate of  electron energy losses,
$D(r)$ is the coefficient of spatial diffusion, and the function
$Q$ describes the injection spectrum and spatial distribution of
sources.

\begin{figure}[ht]
\epsscale{.80} \plotone{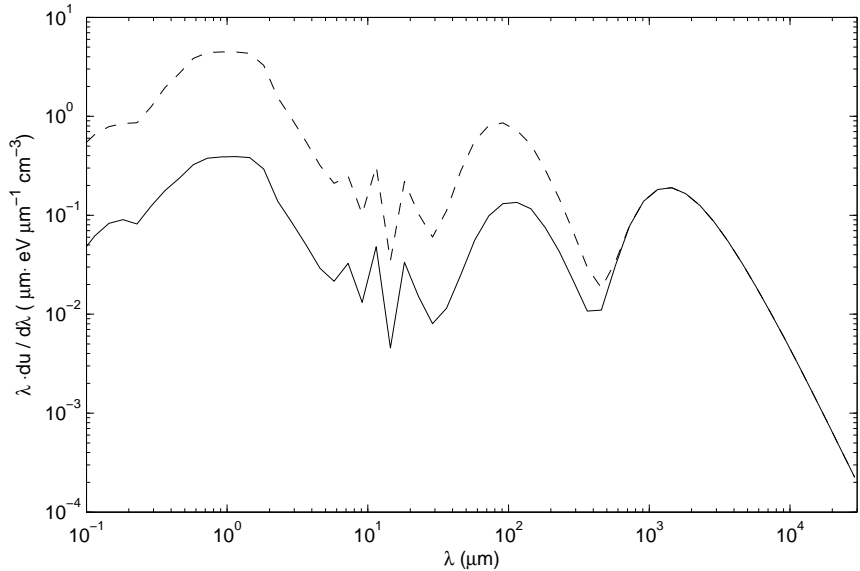} \caption{Spectrum of background
photons at the 47 Tuc (solid line) and at Terzan 5 (dashed line).\label{photon}}
\end{figure}

Relativistic electrons loose their energy by interacting with the
interstellar magnetic field (synchrotron losses) and with
background photons (inverse Compton losses). The strength of
magnetic field in the interstellar medium is about 3 $\mu$G. There
are three components of background photons in the Galaxy which
interact with electrons: they are relic, infrared and optical
photons. In Fig. \ref{photon} we present the spectra of background
photons at the position of 47 Tuc and Terzan 5 in the Galaxy which
were obtained with the GALPROP code (Strong \& Moskalenko 1998). However inside
GCs we have an additional component of optical photons which are
emitted by stars of the cluster. Their spatial distribution is
strongly nonuniform. It may reach the value about $w^0_{op}=300$
eV cm$^{-3}$ for 47 Tuc and about $w^0_{op}=100$ eV cm$^{-3}$ for
Terzan 5 in the cluster center but it decreases rapidly with the
distance from the GC center. Thus, for 47 Tuc the spatial
distribution of optical photons was derived by Michie (1963) and Kuranov \& Postnov (2006)
which is:
\begin{equation}
w_{op}(r) = w^0_{op} \times \left\{
\begin{array}{cc}
    1&\mbox{, for }r<r_c\\
    \left(r_c/r\right)^2&\mbox{, for }r_t>r>r_c\\
    (r_c r_h)^2/r^4&\mbox{, for }r>r_t
\end{array}\right.
\label{r_ph}
\end{equation}
where $r_c = $ 0.5 pc, $r_t = 50$ pc and $r_h = \sqrt{2r_c r_t/3}$.

The diffusion coefficient inside clusters ($r<r_c$) is supposed to
be smaller than in the surrounding medium, that seems to be reasonable since GCs
have high densities of stars with their winds which can create turbulence in the medium
inside GCs. Therefore, the diffusion
coefficient was taken in the form
\begin{equation}
D(r)=D_0 + D_1\theta(r-r_D)
\end{equation}
where $\theta(r)$ is the Heaviside (step) function. The values of $D_0$  and $D_1$ are derived from the data
(see below). The value of $r_D = 100~\mbox{pc} \simeq 2r_c$ is also derived from the spatial distribution of
gamma-ray emission.

We assume that pulsars inject a monoenergetic spectrum of
electrons in the form
\begin{equation}
Q(r,E)=\sum\limits_n\frac{L_{sd}^n}{E_{inj}^n}\delta(E-E_{inj}^n)\delta(r-
r_n),
\end{equation}
where $r_n$ is the position of  the $n^{th}$ pulsar, and the injection process is stationary.  Here $L_{sd}^n$ is the
spin-down loss rate of the $n^{th}$ pulsar in the globular cluster
and the injection energy of electrons generated by each pulsar
is estimated as (see Eq. (\ref{ginj}))
\begin{equation}
E_{inj}=10^2f_{e^{\pm}}^{-1}L_{34}^{1/2}{\rm GeV} =E_0L_{34}^{1/2}
\label{inj}
\end{equation}
where $E_0$ is a constant.  In section 3 we have pointed out that $f_{e^{\pm}}\sim 1$ for MSPs without outer gaps and $f_{e^{\pm}}\sim 30$ for MSPs with outer gap, $E_0$ should be either $\sim 10^2$GeV or $\sim 5$GeV. If the mean spin-down power of MSPs $L_{34}\sim 2$ then the most possible range of $E_{inj}$
should be 200GeV for MSPs without outergaps and 10GeV with outergap.

For the source function $Q$  we estimate the cumulative
contribution of all pulsars in the cluster. The required number of pulsars we
estimate from the observed intensity of GeV gamma-ray emission
from the clusters.

The process of IC scattering depends on the parameter
$\xi=m_ec^2/\epsilon \gamma$, where $\epsilon$ is the energy of a
background photon and $\gamma$ is the gamma-factor of electrons.
If $\xi>1$ than the scattering is classical and the total
cross-section of IC scattering equals the Thompson cross-section,
$\sigma_T$. In the case of $\xi<1$ the cross-section drops down as
$\sigma\propto \sigma_T/\gamma$.

 We notice that
scattering of relativistic electrons on relic and on
IR photons satisfies the condition $m_ec^2>\epsilon \gamma$ and
therefore is classical. For interactions of these electrons
with optical photons this condition may be violated. Then the
interaction photon-electron is catastrophic when a significant
part of the electron energy is transferred to a scattered photon.
In this case the exact Klein-Nishina cross-section is used for
calculations.

 The scattered photon spectrum per electron is (see Blumenthal \& Gould 1970):
\begin{equation}\label{KN_cross}
\frac{d^2N}{dtdE_1} = \frac{2\pi r_0^2mc^3}{\gamma}\int
\frac{n(\epsilon)d\epsilon}{\epsilon} \times \left[ 2q\ln q +
(1+2q)(1-q) + \frac{1}{2}\frac{(\Gamma q)^2}{1+\Gamma
q}(1-q)\right] \,
\end{equation}
where $\epsilon_1$ is the energy of scattered photon, $E_1 =
\epsilon_1/\gamma mc^2$, $n(\epsilon)$ is the density of
background photons, $\Gamma = 4\epsilon\gamma/mc^2$, $q = E_1/
\Gamma (1-E_1)$. The range of values of $E_1$ is restricted by the
range
\begin{equation}\label{E1_limits}
1 \gg \epsilon/\gamma mc^2 \geq E_1 \geq \Gamma/(1+\Gamma)
\end{equation}

The rates of electron energy losses  in the two limit cases (the
classical (Thompson) limit  and the extreme Klein-Nishina limit)
are:
\begin{equation}
\left(\frac{dE}{dt}\right)_T = \frac{4}{3}\sigma_T c\gamma^2w_{em}~~~
\left(\frac{dE}{dt}\right)_{KN} = \pi r_0^2 m^2c^5 \int \frac{n(\epsilon)d\epsilon}{\epsilon}
\left(\ln\frac{4\epsilon\gamma}{mc^2} - \frac{11}{6}\right) \,
\end{equation}
where $w_{em}$ is the energy density of  background photons.

To compare with observational data we calculate the two
parameters of  IC gamma-ray flux from GCs. The first one is
the flux of gamma-ray emission from the cluster
\begin{equation}
F(\epsilon_1) = \int dr\int dE f(r, E) \frac{1}{\gamma mc^2}
\frac{d^2N}{dtdE_1}\label{spec_e}
\end{equation}
where $f(r,E)$ is the solution of Eq. (\ref{pr_state}).

\begin{figure}[ht]
\epsscale{.50} \plotone{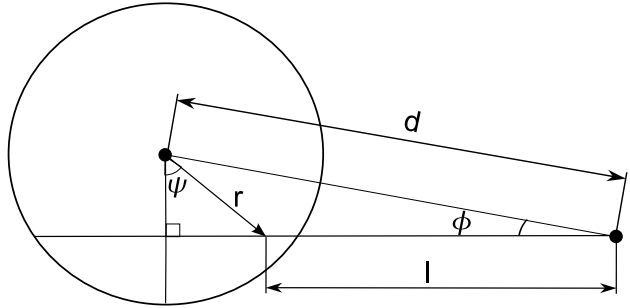} \caption{The schematic view of the cluster from Earth.}\label{schm}
\end{figure}

The second is the spatial distribution of the IC flux in the
energy range $\Delta\epsilon$ as observed from Earth:
 \begin{equation}
\Phi(\phi) = \int\limits_{\Delta\epsilon}d\epsilon_1\int
f\left(r, \epsilon_1\right)dl( \phi) \label{spec_r}
\end{equation}
Here $l$ is the line of sight and $\phi$ is the angular distance
from the center as observed from Earth. The distance from the center of the cluster $r$ can
be estimated in the following way (see Fig. \ref{schm}):
\begin{equation}
r(l,\phi) = d\cdot \frac{\sin \phi}{\cos \psi}\mbox{.}
\end{equation}
The Eq. (\ref{spec_r}) may be rewritten as
\begin{equation}
\Phi(\phi) = \int\limits_{\Delta\epsilon}d\epsilon_1\int\limits_{-\pi/2}^{\pi/2-\phi}
f\left(r, \epsilon_1\right)\frac{dl}{d\psi}d\psi \label{spec_r_angular}
\end{equation}
where
\begin{equation}
\frac{dl}{d\psi} = d\cdot\frac{\sin \phi}{\cos^2 \psi}\mbox{.}
\end{equation}

The energy of primary electrons $E$ and scattered gamma-ray
photons $\epsilon_1$ are related with each other as:
\begin{equation}\label{IC_ph_en}
\epsilon_1=\frac{4}{3}\epsilon\left(\frac{E}{mc^2}\right)^2
\end{equation}
in the classical limit, and $\epsilon_1\simeq E$ in the extreme Klein-Nishina limit.

The actual spin-down rate of individual MSPs in globular clusters is very difficult to be determined due the very strong gravitational force
in the core of globular cluster. Some attempts have been made to subtract the effect of the gravitational effect and recover the
true spin-down rate of MSPs in globular clusters (e.g. Freire et al. 2001;Grindlay et al. 2002).
In general such subtraction scheme is very reasonable, however it may not be correct for individual pulsar. For example the average
gravitational field used is a simple function of distance from the center of the cluster to the pulsar. The observation can only determine the
project distance instead of the actual distance. Therefore it is questionable whether the estimated spin-down rate for individual pulsar
is correct. On the other hand this method may provide a good correction on average. In this paper, we assume for simplicity that each MSP in globular cluster has the same spin-down power and we use the average
spin-down power $\sim 2\times 10^{34}$erg/s estimated by
Freire et al. (2001) and Grindlay et al. 2002 as the characteristic spin-down power of each MSP in our subsequent calculation.
Then the total injection spectrum of electron/positron pairs produced by all pulsars of the cluster can be assumed as monoenergetic,
$Q(E)\propto \delta (E-E_{inj})$, and the total number of electron with energy $E$ in the cluster is evaluated under the influence of IC/synchrotron
losses that gives
\begin{equation}
\frac{dN_e}{dE_e}=f(E_e) \propto E_e^{-2}\theta (E_{inj} -E_e),
\end{equation}
where $E_{inj}$ is given by equation (9).

\section{Applications}
\subsection{47 Tucanae}
\begin{figure}[ht]
 \plotone{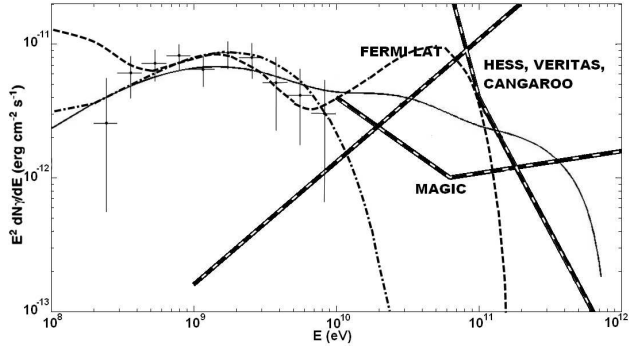} \caption{Gamma-ray flux
from 47 Tuc obtained from the inverse Compton model. The data
point from Abdo et al. (2009). Solid line correspond to relic photon
scattering, dashed line correspond to IR photon scattering and
dash-dotted line correspond to optical photon scattering.
Sensitivities of different gamma-ray instruments are shown by the
heavy dashed lines.}\label{gamma_IC}
\end{figure}

The flux of gamma-rays expected in the IC model for the emission
region within the diameter of 1\degr  around the cluster center
(Eq.(\ref{spec_e})) and the Fermi observational data are shown in
Fig. \ref{gamma_IC}. We find that the optical photons emitted from the core of GC (see Eq. (6)) does not contribute significantly
in the gamma-ray flux produced by the inverse Compton scattering because their density decreases rapidly away from the core and the
diffusion mean free path (see Eq.(20) below) is much larger than the size of the core.
One can see from this figure that the data can equally
be interpreted either by scattering on relic (solid line),
IR (dashed line) or optical photons (dash-dotted line). The three peaks on
each of these lines correspond to scattering on relic, IR, and
optical photons, respectively. Since the characteristic energies of soft photons from relic, IR and optical components are different to make them to be scattered to GeV range one should use electrons with different energies in accordance with Eq. (\ref{IC_ph_en}). The energy parameter from Eq.
(\ref{inj}) corresponding to the relic scattering to be responsible for explanation of FERMI data is $E_{inj}^{relic}
= 0.7$ TeV, to the IR
scattering is $E_{inj}^{IR} = 0.15$ TeV, and to the optical scattering is  $E_{inj}^{op} = 0.02$ TeV.
As one can see from this figure
LAT, MAGIC and even HESS are able to detect the predicted
excesses in the energy range above 10 GeV except the case when the GeV gamma-ray emission is produced by scattering on optical photons (the dash-dotted line).
However these excesses depend on the nature of the soft photons. In general IR photons give the strongest excess in 30 GeV range. If this flux level is detected, it supports that the GeV gamma-rays have IC origin. On the other hand, if the excess is found but significantly weaker than the predicted level of IC model. Then part of GeV gamma-rays may still come from CR mechanism as predicted by Venter \& De Jager (2009). We also want to remark that although the upward scattered relic photons can fit the Fermi data, it requires $E_{inj}$ larger than the estimated value by a factor of 3 (cf. discussion below Eq.(9)), which is unfavorable unless the pair creation process of MSPs in 47 Tuc is strongly suppressed.

The recent HESS observations gave an upper limit $\sim 6.7\times
10^{-13}$ cm$^{-2}$s$^{-1}$ of  gamma-rays photon flux for energies
above 800 GeV (Aharonian et al. 2009) that is higher than we predict
for  47 Tuc.

\begin{figure}[ht]
 \plotone{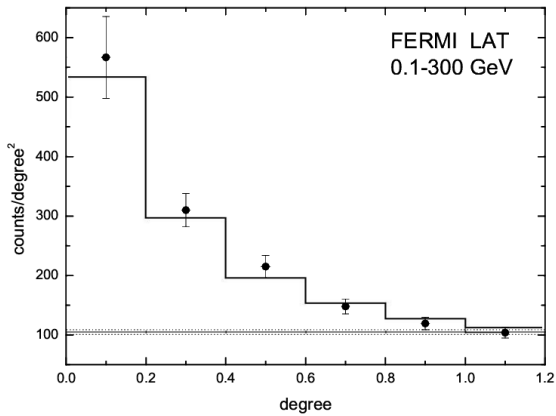} \caption{The spatial distribution of
gamma-ray emission from direction of 47 Tuc and the results of
simulation.}\label{Tuc_r}
\end{figure}

We can compare the spatial distribution as expected from the inverse
Compton model with the observed data. To obtain the $\gamma-$ray brightness
profile of 47~Tuc, we have taken the LAT data obtained from  4 August 2008 to 4
December 2009. For the data filtering, we adopt the standard procedures
suggested by the Fermi Science Support Center (for further details, please
refer to Kong et al. 2010). We have binned the filtered event list into an
image centered at the peak of the $\gamma-$ray emission with a bin size of
$0.1^{\circ}$. The surface brightness profile of the $\gamma-$rays from 47~Tuc
which is displayed in Fig.4. The average background level and its
$1\sigma$ deviation are indicated by horizontal lines, which were calculated by
sampling the source-free regions around 47~Tuc within a $10\times10$
degree$^{2}$ field-of-view. The average background is estimated to have a level
of $105\pm4$~cts~degree$^{-2}$.
The observed data are nicely
reproduced if the diffusion coefficients are the
following (see Eq.(7)): for relic scattering $D_0^{relic} = 10^{27}$ cm$^2$/s,
$D_1^{relic} = 5\times 10^{27}$ cm$^2$/s. For IR scattering
$D_0^{IR} = 6\times10^{26}$ cm$^2$/s, $D_1^{IR} = 6\times
10^{28}$ cm$^2$/s. For optical scattering $D_0^{op} = 10^{26}$ cm$^2$/s,
$D_1^{op} = 10^{27}$ cm$^2$/s. However we want to emphasize that the contribution of unresolved point-like sources in
the total gamma-ray flux of 47 Tuc observed by $Fermi$ is unknown. If future observations show that this flux is really diffuse,
it proves its IC origin.  Here we have assumed that most point sources are located inside the core of globular clusters and
the angular resolution of future observations are good enough to remove the contribution from the core.

\subsection{Terzan  5}
\begin{figure}[ht]
 \plotone{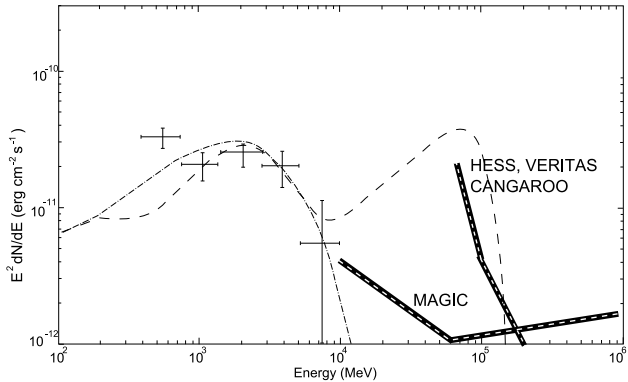} \caption{Gamma-ray flux
from Terzan 5 obtained from the inverse Compton model. Dashed line corresponds to IR photon scattering and
dash-dotted line corresponds to optical photon scattering.}\label{Ter}
\end{figure}

The expected flux of gamma-rays from Terzan 5 and the Fermi data are
shown in Fig.  \ref{Ter} by using the same set of diffusion coefficients as for 47 Tuc and the injected energy equals 180GeV for IR photons and 25 GeV for Optical photons.  We find that it is impossible to use the relic photons to obtain reasonable fit to the Fermi data and the scattering on relic photons provides
a negligible effect  because of very high density of IR and optical photons in
Terzan 5. On the other hand, the IR or optical scattering  can nicely reproduce the
experimental data as shown in Fig. \ref{Ter}.

As Terzan~5 is located in a more complicated environment than 47~Tuc, in
particular it is located very close to the Galactic plane (see Figure~1 in Kong
et al. 2010), this makes the estimation of its $\gamma-$ray brightness profile
much more intricated. In view of this difficulty, we do not compare the
spatial distribution computed from the model with the observation for Terzan~5.

\subsection{Model fitting of other globular clusters}
\begin{figure}[ht]
 \plotone{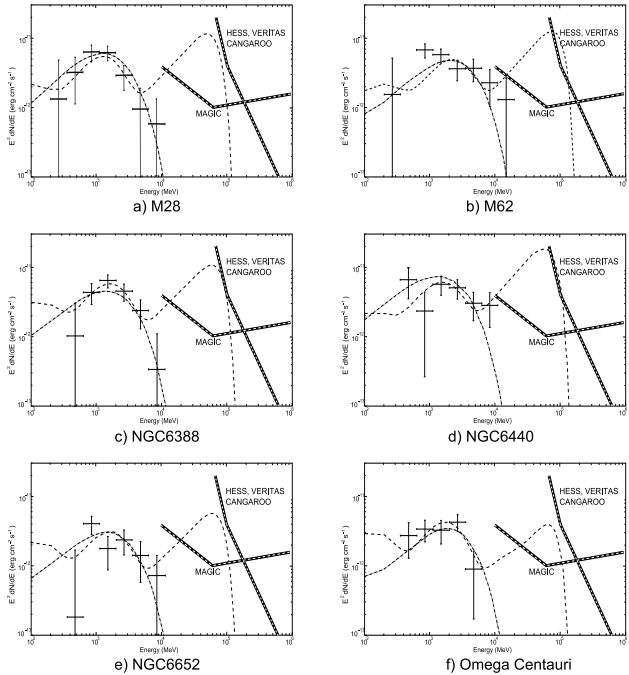}
 \caption{Experimental data for six globular clusters by Abdo et al. (2010b) along with IC model data for IR photons scattering (dashed line)
 and optical photons scattering (dash dotted line).}
 \label{gc1}
\end{figure}

We also apply our IC model to other clusters presented in the paper by Abdo et al. (2010b). The diffusion coefficients are the same as for 47 Tucanae and Terzan 5 for simplicity. Although the ICS of the relic photons in some GCs may also fit the Fermi data, as we have pointed out in section 2.2 without another factor the millisecond pulsar ratio between Terzan-5 and 47 Tuc is 12 required by gamma-ray luminosity ratio, which seems too large in comparing with the observed millisecond pulsar ratio $\sim 1.5$. Since the relic photon density is constant everywhere, it cannot reduce this ratio.
Furhtermore they also cannot fit the Fermi data for Terzan-5 and the required $E_{inj}$ is higher than the theoretical predicted value for 47 Tuc, therefore we conclude that the relic photons may not be the possible background soft photons to produce the gamma-rays in Fermi energy range. We will not use them in fitting the spectrum of the other six globular clusters in this subsection. However the relic photons can still participate in the IC process and they can contribute to X-rays significantly.
In fitting these six new globular clusters we vary slightly the parameter $E_{inj}$ for different clusters. The values of $E_{inj}$ as well as effective output power $\eta L_{sd}$, which is treated as normalization factor in fitting, are presented in Table \ref{gc1_table}, which are fixed by comparing with the observed gamma-ray power of each cluster.
The spectra of the clusters together with the data point from Abdo et al. (2010b) are presented in Fig. \ref{gc1}.
We can see that the IC model can fit the gamma-ray spectra of all eight globular clusters with similar parameters well (cf. Fig. \ref{gc1} and Table \ref{gc1_table}). In estimating the gamma-ray power from the globular clusters we have used the observed energy fluxes and distances given in Abdo et al. (2010b).

\begin{table}[ht]
	\centering\caption{Fitting parameters for IC model for 8 clusters from Abdo et al. (2010b)}\label{gc1_table}
		\begin{tabular}{| l || c  c || c  c |}
		\hline
		Name & Infra-Red & & Optical & \\
		\hline
		     & $E_{inj}$ (GeV) & $\eta L_{sd}~10^{34}$ erg/s & $E_{inj}$ (GeV) & $\eta L_{sd}~(10^{34}$ erg/s)\\
		\hline
		M28 & 130 & 14.8 & 17 & 6.2\\
		M62 & 180 & 21.8 & 25 & 10.9\\
		NGC 6388 & 150 & 51.6 & 20 & 25.8\\
		NGC 6440 & 150 & 47.5 & 20 & 19.0\\
		NGC 6652 & 150 & 20.6 & 20 & 7.8\\
		Omega Centauri & 150 & 6.1 & 20 & 2.8\\
		Terzan 5 & 180 & 49.1 & 25 & 25.7\\
		47 Tucanae & 150 & 10.0 & 20 & 4.8\\
			\hline
			
		\end{tabular}
\end{table}

\subsection{Implications of the fitting parameters}

In general we need three parameters to fit the observed spectrum of globular clusters, i.e. diffusion coefficient, injected energy ($E_{inj}$) and $\eta$. Since we are fitting the total spectrum instead of the spatial dependent spectrum then the diffusion coefficient mainly controls the size of emission region, for simplicity we have assumed that the diffusion coefficients of other globular clusters are the same as those of 47 Tuc. The exact value of the diffusion coefficient must be determined by measuring the angular size of the diffusion emission region. In IC model we predict that a very wide energy bands will be produced (cf. next section), therefore the angular size of other energy bands, e.g. radio and X-rays, can also be used to estimate the diffusion coefficient. The injected energy of pairs controls the spectral break and finally $\eta$, which
can be interpreted as the efficiency for conversion of spin-down power to gamma-ray power times the total number of millisecond pulsars in the cluster, controls the magnitude of spectrum.

The injected energy given by Eq. 9 depends on $f_{e^{\pm}}$ (cf. Eq. 2). In section 3 we discuss the possible values of $f_{e^{\pm}}$. If outergap exists in MSPs, $f_{e^{\pm}}\sim 30$, which gives $E_{inj}\sim 10 GeV$, which is less than the fitting values by a factor of 2 for Optical photons. On the other hand, if pairs are only produced in polar gap, we have estimated in section 3 that $f_{e^{\pm}}\sim 1$, which gives $E_{inj}\sim 2\times 10^2 GeV$. This estimate is consistent with the fitting values for IR as soft photons. However, if the optical photons are the soft photons then this estimate is higher than the fitting values by a factor of 10 in general. This may imply outergap exist but its pair production multiplicity is substantially lower than previous model estimates, for example instead of closing the outergap in terms of photon-photon creation process outer gap closed by magnetic pair creation is possible (cf. Takata et al. 2010), which gives less outgoing pairs. However, if this is the case, CR contribution can not avoided. The observed gamma-rays in Fermi energy range should be a mixture of CR and IC processes.

In fitting data point of view, $\eta L_{sd}$ is the normalization factor. In our model it can be estimated by relating the observed gamma-ray power $L_{\gamma}$ to the theoretical IC power, i.e. $N_{MSP}L_{e^{\pm}}$, where $N_{MSP}$ is the total number of MSPs in the globular cluster and $L_{e^{\pm}}$ is the part of spin-down power carried away by the pairs given by Eq. 4. If we assume that each pulsar has similar spin-down power, e.g. $L_{34}\sim 2$, once $f_{e^{\pm}}$ is fixed then we can use the above conservation to estimate the total number of MSPs in the globular cluster. Let's assume the outer gap does not exist, we have estimated in section 3 that the fraction of spin-down power carried away by pairs is about 0.1. Using table 1 and assuming IR as the inverse Compton soft photons, we can estimate that the number of MSPs for 47 Tuc and Terzan-5 are $\sim$50 and $\sim$245 respectively.


\section{Model Constraints by other energy bands}
Although the inverse Compton scattering can explain the Fermi data of both clusters very well,
we cannot distinguish from the data scattering on which photons, i.e. optical, IR and relic, produce this gamma-ray flux.
For 47 Tuc all three cases are equally possible. For  Terzan 5 the scattering on galactic infrared photons and optical photons can be possible candidates. In this section we will explore the constraints for the model in derived from other energy bands.

The inverse Compton scattering cooling time is given by $\tau_{cooling}\sim 4\times 10^{14} \gamma_{w5}^{-1}w_{-12}^{-1}\rm s$, where  $\gamma_{w5}$ is the Lorentz factor of the relativistic electron/positron pairs in units of $10^5$ and $w_{-12}$ is the energy density of soft photon in units of $10^{-12} \rm erg/cm^3$. The diffusion time of these pairs over the distance $d$ is given by
$\tau_d \sim 10^{11} d^2 D_{26}^{-1} \rm s$, where $d$ is in units of pc and $D_{26}$ is the diffusion coefficient in units of $10^{26}\rm cm^2/s$. Therefore the diffusion radius is estimated as from the equality $\tau_{cooling}=\tau_{d}$ and is given by
\begin{equation}
d\approx 63 \gamma_{w5}^{-1/2}w_{-12}^{-1/2}D_{26}^{1/2}\rm pc.
\end{equation}

Since the total IC photon spectrum from the GC is given by
 \begin{equation}
 \Phi(\epsilon_\gamma)=\int\limits_{E_e}n_{ph}(\epsilon_{ph})c
 \frac{dN}{dE_e}\frac{d\sigma_{IC}}{d\epsilon_\gamma}dE_e,
 \end{equation}
where $\frac{dN}{dE_e}$ is given by Eq.(19), therefore the photon spectral index is $\sim -1.5$
(see Blumenthal \& Gould 1970).
Here $d\sigma_{IC}/d\epsilon_\gamma$ is the IC differential
cross-section which in the classical limit is approximately
\begin{equation}
\frac{d\sigma_{IC}}{d\epsilon_\gamma}=\sigma_T
\delta\left(\epsilon_\gamma-
\frac{4}{3}\epsilon_{ph}\left(\frac{E_e}{mc^2}\right)^2\right) ,
\end{equation}
The energies $\epsilon_{ph}$ and $\epsilon_\gamma$ are the energies of background
and scattered photons respectively, and $n_{ph}$ is the  photon density of background photons.

The power in IC X-ray emission with the energy $\epsilon_x$ can be produced by scattering on different background photons. As compared with the contribution from scattering on the relic photons we have
\begin{equation}
\frac{\Phi(\epsilon_x)_{relic}}{\Phi(\epsilon_x)_{ph}}\simeq\frac{w_{relic}}{w_{ph}}
\sqrt{\frac{\epsilon_{ph}}{\epsilon_{relic}}}
\end{equation}
where $\epsilon_{relic}$ and $\epsilon_{ph}$ are the energies of relic and any other
sort of background photons, and $w$ is the corresponding energy density of photons.
Eq. (23) can be directly obtained by integrating Eq. (21) subject to the constraint of the $\delta$-function of Eq. (22), i.e. after integration using $E_e = mc^2(\epsilon_x/\epsilon_{ph})^{1/2}$ to replace $E_e$, and $w_{ph}=\epsilon_{ph}n_{ph}$.
In Fig. 1 we can see that for 47 Tuc the energy densities of different soft photons are very closed, therefore the IC X-ray emission is mainly contributed from the inverse Compton scattering of relic photons. For Terzan 5, although the ratios of the energy density  between the IR photons and the relic photons, and between the optical photons and the relic
photons are $\sim 4$ and $\sim 40$ respectively, $\sqrt{\frac{\epsilon_{IR}}{\epsilon_{relic}}}$ and $\sqrt{\frac{\epsilon_{optical}}{\epsilon_{relic}}}$ are close to $\sim 4$ and $\sim 40$ respectively. Therefore the contributions to IC X-rays by the relic photons, IR and optical photons are comparable. However it is very important to note that although the energy flux of IC X-rays from each of these three soft photons is comparable, they are emitted from different region. For example most of IC X-rays by scattering relic photons come from a few arcmins region whereas IC X-rays by scattering IR photons and Optical photons come from much bigger radius because the electrons/positrons are cooling off on the way diffusing out from the core.

The X-ray energy flux at 5 KeV ($F(\epsilon_{\gamma}=5KeV)$) can be estimated as follows. Since IC energy flux is given by $F(\epsilon_{\gamma})\approx \epsilon_{\gamma}^2\Phi(\epsilon_{\gamma})\sim \epsilon_{\gamma}^{1/2}$ because $\frac{dN}{dE_e}\sim E_e^{-2}$,
we can estimate the energy flux at 5 KeV $F(\epsilon_{\gamma}=5KeV)$ from its peak energy flux. We have argued that the relic photons should be
the most important photons to generate the X-rays through IC process, therefore $F(\epsilon_{\gamma}=5KeV)$ produced by IC of relic photons
is given by
\begin{equation}
F_x(5KeV) \approx (5KeV/8\gamma_{w5}^2MeV)^{1/2}F_{relic},
\end{equation}
where $F_{relic}$ is the peak energy flux of the inverse Compton scattering relic photons and the characteristic upward scattering energy of relic photons is $\sim 8\gamma_{w5}^2MeV$. We can estimate the peak energy flux of the scattered relic photons by $F_{relic}=(w_{relic}/w_{soft})F_{\gamma}^{obs}$, where $F_{\gamma}^{obs}$ is the observed gamma-ray energy flux in the GeV energy range, $w_{relic}$ and $w_{soft}$ are the energy density of the relic photons and the soft photons, which upward scatter to produce gamma-rays.

The strength of magnetic field $B$ near the clusters is not known exactly, it is estimated to be of order of $10^{-6}$G (Beck et
al. 2003).
The energy loss ratio between synchrotron radiation and inverse Compton scattering is given by
\begin{equation}
F_{syn} \approx F_{\gamma}\frac{B^2/8\pi}{w^{soft}} = 3\times 10^{-2} B_{-6}^2 (w^{soft}_{-12})^{-1} F_{\gamma},
\end{equation}
where $B_{-6}$ is the magnetic field in units of $10^{-6}$G. We can see that the synchrotron loss is not negligible. The characteristic synchrotron frequency is given by
\begin{equation}
\nu_{syn} = \gamma_w^2\frac{eB}{2\pi mc} = 4.4 \times 10^{10}\gamma_{w5}^2 B_{-6} \rm Hz,
\end{equation}
which is in the radio band. We can estimate the energy flux at 1GHz
\begin{equation}
F_{1GHz} \approx (1GHz/\nu_{syn})^{0.5}F_{syn}
\end{equation}
if $\nu_{syn}> 1GHz$.  $F(\nu)$ corresponds to energy flux but it is more useful to estimate the differential flux per Hz measured in Jy. To obtain it one can divide the
energy flux by characteristic frequency.

\subsection{47 Tuc}
We pointed out that the angular resolution of $Fermi$ is of order of $\sim 1^{\circ}$ and the angular resolution of HESS is also of order of $\sim 1^{\circ}$. This angular size implies that the emission radius of 47 Tuc is $\leq$80pc. In fitting the gamma-ray spectrum we find that basically we cannot differentiate which kind of soft photons produces the observed gamma-rays. However from Eq.(24) the predicted X-ray energy flux depends on the Lorentz factors, which are $\gamma_w = 1.4\times 10^6$ for relic photons, $\gamma_w = 2.8\times 10^5$ for IR photons and $\gamma_w = 4\times 10^4$ for optical photons respectively. The predicted X-ray energy fluxes are
$F_x(5KeV) \approx (5KeV/8\gamma_{w5}^2MeV)^{0.5}(w_{relic}/w_{soft}) F_{\gamma} \sim 1.8 \times 10^{-14}\rm erg~ cm^{-2}~s^{-1}$
for relic photons, $\sim 10^{-13}\rm erg ~cm^{-2}s^{-1}$ for IR photons and $\sim 3.2 \times 10^{-13}\rm erg~ cm^{-2}s^{-1}$ for optical
photons respectively in 1 degree radius.

With the Chandra observation, Okada et al. (2007) have reported two extended
X-ray features potentially associated with 47 Tuc which are labeled as T1 and
T2 in their Fig. 1a. However, a recent deep Suzaku observation reported by
Yuasa et al. (2009) found the X-ray spectrum T1 is consistent with a
red-shifted thermal plasma and suggest its nature as a background galaxy
cluster. On the other hand, T2 is relatively fainter and locates just outside
the half-mass radius of 47 Tuc. Its spectrum can be modeled by a power-law with
a photon index of $\Gamma\sim2.2$. The flux of this feature is found to be
$\sim7\times10^{-14}$ erg cm$^{2}$ s$^{-1}$. Although the interpretation that
this feature arises via ICS is tempting (see also Krockenberger \& Grindlay
1995), it should be noted that it locates very close to the very bright
emission of T1. Also both features locate at a large off-axis angle in this
Chandra observation which result in a rather wide point spread function at
their locations. Therefore, at least a fraction of the X-rays from T2 can
possibly be contributed by T1.  Furthermore, the tidal radius of 47 Tuc is 43'  and it is possible that a good fraction of millisecond pulsars are located outside the half-mass radius but within the tidal radius. Consequently the center of this extended faint X-ray source T2 may not coincide with the half-mass radius.
With this consideration, the flux measured from
T2 should be considered as an upper limit. The largest model predicted X-ray
energy flux in 3' radius resulting from optical photons is $\sim
(3'/1^{\circ})^2 3.2 \times 10^{-13}\rm erg~cm^{-2}s^{-1} \sim 10^{-15}\rm erg
~cm^{-2}s^{-1}$. However, it is very important to note that the actual emission
region of gamma-ray can be much smaller than $1^{\circ}$ as this estimate is
limited by the angular resolution of LAT. Therefore, a dedicated X-ray
observation with T2 on-axis can provide important constraint for the model
parameters.


According Eq. (27), the energy fluxes at 1 GHz are $9\times 10^{-14} \rm erg ~cm^{-2} s^{-1}$ for optical photons, $5\times 10^{-14} \rm erg ~cm^{-2} s^{-1}$ for IR photons and $5\times 10^{-15} \rm erg ~cm^{-2} s^{-1}$ for relic photons in  $1^{\circ}$, which correspond to 9 Jy, 5 Jy, and 0.5 Jy respectively. At 400 MHz the corresponding fluxes will be equal to 18 Jy, 8 Jy and 0.7 Jy.

The radio flux from region with diameter 1$^\circ$ is 19 Jy at 408 MHz (Haslam et al. (1982)) and 27 Jy at 1420 MHz (Reich et al. (2001)).
However, in view of the poor resolution of the instrument, there may have contamination by other sources. Therefore, the true radio fluxes due to the pulsar wind at these frequencies should be lower than the aforementioned values. In view of this, these observed values should only be considered as the upper limits. Since the theoretical estimate at 400 MHz for the background optical photons (i.e. 18 Jy) is comparable with the observational limit reported by Haslam et al. (1982), there is a high probability that the IC model with the optical photons as the soft photon field may over-predict the radio flux. In section 5.4, we have pointed out that the injected energy $E_{inj}$ for optical photons is less than the model predicted value by a factor of $\sim 10$ if the outer gap does not exist (cf. Eq. 9 and Table 1). If the outer gap indeed exists in MSPs of Globular clusters, CR must contribute to GeV gamma-rays and hence the contribution by the IC component is only partial. The predicted diffuse radio flux for optical above is assumed all observed GeV gamma-rays result from IC. If this is not the case then the reduction of the diffuse radio flux should be prorata.
On the other hand, for the other soft photon fields (i.e. IR and relic photons), the IC model-predicted flux densities appear to be more consistent with this limit. At 1 GHz, the IC model-predicted values for all the soft photon fields in our consideration are far below the observed upper bound at 1.4 GHz (Reich et al. 2001). This suggests that the currently available observational results do not allow us to put a tight constraint at this frequency. Future radio observations with higher resolution and sensitivity can possibly help us to discriminate different scenarios. Since the radio flux and gamma-ray flux are correlated, the more detail observations in radio band may provide better constraint on these models.

\subsection{Terzan 5}
{\it Chandra} has also detected diffuse X-Ray emission in 2-7 keV band from Terzan 5  (Eger,
Domainko \& Clapson 2010) and the X-ray energy flux is
$5\times 10^{-13}$ erg  cm$^{-2}$s$^{-1}$. Unlike in the case of 47 Tuc, whose diffuse X-ray is most likely from the unresolved X-ray point sources as suggested by Okada (2005), some diffuse X-ray emission in Terzan 5 clearly exists from 90" to 160" even an unresolved point-source is subtracted. In other words the X-ray emission region is $\sim$10pc. If the diffuse X-rays is the tail of the inverse Compton scattering, we can use it to constrain the theoretical models. In fitting the gamma-ray spectrum of Terzan 5, either upward scattering the galactic infrared photons or optical photons are possible. However the required Lorentz factors for IR and Optical are $2.8\times 10^5$ and $4\times10^4$ respectively, and the energy density are $\sim 10^{-12}\rm erg ~cm^{-3}$ and $\sim 6\times 10^{-12}\rm erg ~cm^{-3}$ respectively. If the emission region is really 10pc, then the diffusion coefficient of Terzan 5 is much smaller than that of 47 Tuc. According to Eq.(20) it gives $D\sim 10^{25}cm^2 s^{-1}$. The locations of 47 Tuc and Terzan 5 are very much different, the former is above the galactic plane and the latter is in the galactic plane. This factor may cause the difference in the diffusion coefficient. According to Fig. 1, $\epsilon_{relic}/\epsilon_{IR}\sim 0.3$ and $\epsilon_{relic}/\epsilon_{optical}\sim 0.05$, by using Eq.(24) the predicted X-ray energy fluxes $\sim 7\times 10^{-14}\rm erg ~cm^{-2} s^{-1}$ and $\sim 10^{-13}\rm erg ~cm^{-2} s^{-1}$ respectively. These predicted values are about a factor of 3-4 lower than that of the observed value. However in Fig.2 of Eger, Domainko \& Clapson (2010) we can see that if the unresolved X-ray point-source can contribute to the diffuse X-ray, then after subtracting this contribution (the green solid curve) the real diffuse X-ray flux is actually reduced significantly.

Again we can use Eq.(27) to estimate the predicted radio
energy flux, which gives $\sim 5\times 10^{-14}\rm erg ~cm^{-2} s^{-1}$ for the IR model and $\sim 3.4\times 10^{-14}\rm erg ~cm^{-2} s^{-1}$ for the Optical model (5 Jy and 3.4 Jy). We don't have radio data for 3' region around Terzan 5.
If observations show the lower values of radio flux from the corresponding region then radio emission, gamma-ray emission and X-ray emission should occupy
more extended region. In that case the observed diffuse X-ray emission from 3' region should not be related to IC model and should have different nature.
The more detailed observations including spatial and spectral information by Fermi and other higher energy detectors, like HESS, MAGIC, VERITAS etc. as well as radio observations can provide better constraints for the models.

\section{Discussion}

We have calculated GeV gamma-ray spectrum produced by inverse Compton scattering between the relativistic $e^{\pm}$ pairs of the pulsar wind and the background soft photons, which include the relic photons, the star lights of the cluster, the infrared photons and the star light photons from the Galactic disk. We obtain the steady state spatial and energy distribution function of $e^{\pm}$ pairs by solving the standard diffusion equation for describing the cosmic ray propagation  in the interstellar medium. We find that most of high energy radiation comes from region outside the core of globular clusters with a radius $>10$pc. In fact the contribution by upward scattering the star light photons inside the cluster core region is negligible in contradiction to previous calculations (e.g.Bednarek \& Sitarek (2007)). For 47 Tuc the GeV photons detected by Fermi can be reproduced by the upward scattering of all three possible background soft photon fields, i.e. relic photons, IR photons and optical photons. There are no compelling evidence to rule out any of these three models, but the required energy of electrons/positrons for Compton upscattering the relic photons to GeV energy range is a factor of 3 higher than that predicted by the model.
For Terzan 5 both the galactic IR and optical photons are possible soft photons for upward scattering to produce the GeV gamma-rays. Again no compelling evidence to differentiate these two models. Obviously the optical one cannot produce photons higher than 10 GeV.

It is generally agreed that gamma-ray emission from globular clusters are associated with MSPs inside the clusters. It has been standard to explain the spectra of almost all Fermi-LAT pulsars including MSPs, except very young Crab-like pulsars, in terms of CR mechanism, i.e. gamma-rays are emitted from inside the light cylinder. In this paper we propose an alternative model, which fits the GeV-spectra of all 8 Fermi detected globular clusters very well. In IC model it predicts that (1)100MeV-100GeV spectrum is correlated and hence some globular clusters should be sources for MAGIC and HESS. (2)Although IC is the main energy dissipation process, the synchrotron radiation cannot be avoided and it results in diffuse radio emission. This prediction can be best tested by SKA, which has both excellent sensitivity and spatial resolution. (3)The gamma-ray power from globular clusters is not only dependent on the number of MSPs but also depends on the galactic soft photon density at the location of globular clusters, which also a test between CR model and IC model. However, even though all these predictions are correct, this cannot rule out the CR model because it is possible that some fraction of the observed gamma-ray photons are mixture of two origins. Actually it is better to subtract the contribution from the CR model from the data and then compare with the model predictions of IC model. However it is practically impossible to carry out such analysis because in order to calculate the CR spectrum from pulsars accurately the period and magnetic field of each pulsar must be known. Most MSPs in globular clusters have not been found even for 47 Tuc and Terzan-5. The contributions from these undetected MSPs are extremely difficult to estimate. In fact Omega Centauri does not have any reported MSPs. These make such subtraction scheme impossible. Therefore it is more important to prove the predictions of IC model, i.e.  the diffuse emission in various other energy bands, i.e. radio, X-rays and VHE. In future we can use these data to constraint the relative contributions between these two different models.

Finally, we want to remark that by using the 8 GCs reported by Abdo et al. (2010b) and 7 newly confirmed gamma-ray GCs by Tam et al. (2010), Hui et al. (2010b) have carried out a correlation analysis between the
observed $\gamma$-ray luminosities $L_{\gamma}$ and various cluster properties to probe the
origin of the high energy photons from these GCs. They find that $L_{\gamma}$
is positively correlated with the encounter rate $\Gamma_{c}$ and the metallicity $\left[{\rm Fe/H}\right]$, which is an alterative independent estimator for number MSPs in the globular clusters (cf. Hui, Cheng and Taam 2010).
They also find a tendency that $L_{\gamma}$
increases with the energy densities of the soft photon at the cluster location which
favors the scenario that the observed gamma-rays from these GCs
are significantly contributed by the inverse Compton scattering.  It should be noticed that
Hui et al. (2010b)
have used different way to calculate the encounter rate in comparing with Abdo et al. (2010b). Hui et al. (2010b) have included the observed dispersion velocity in evaluating  $\Gamma_{c}$ whereas Abdo et al. (2010b) have used the free fall velocity to approximate the dispersion velocity. For illustration purpose in Figure 7 we
follow the definition of encounter rate given in Abdo et al. (2010b). Figure 7a shows the correlation between the gamma-ray luminosity and the encounter rate, and the correlation coefficient is 0.71. Figure 7b and Figure 7c show the correlations between $L_{\gamma}$ and the combined factor, i.e. $\Gamma_{c}w_{ph}$, where $w_{ph}$ are optical and IR photon energy density respectively. The correlation coefficients for Figure 7b and Figure 7c are 0.79 and 0.82 respectively, which show stronger correlations when the soft photon energy density is included. These results support that the inverse Compton scattering mechanism is at least one of the major gamma-ray emission process in globular clusters.

\begin{figure}[ht]
\epsscale{0.8}
 \plotone{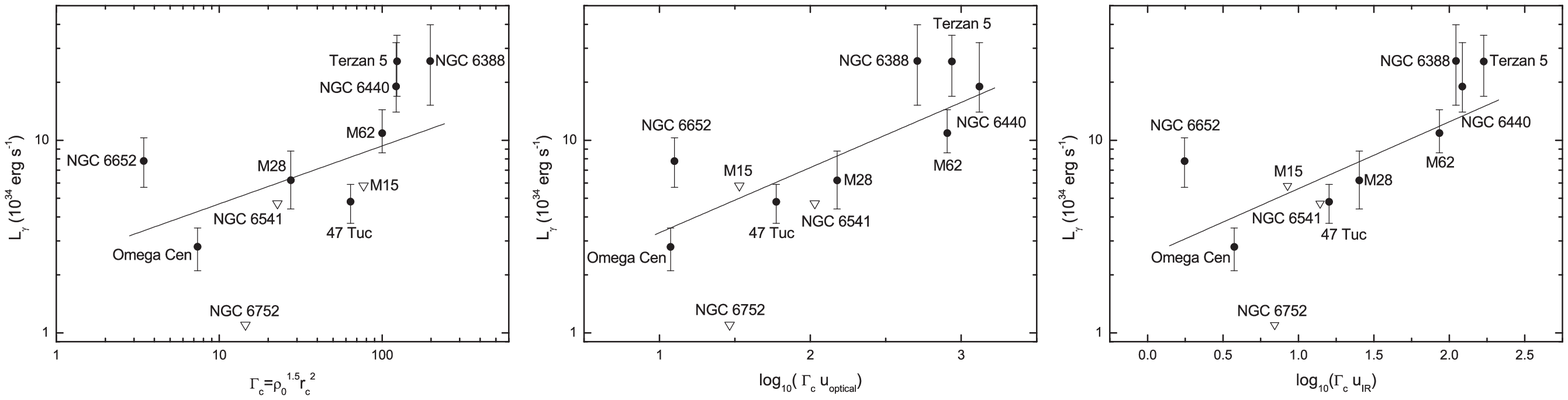} \caption{(a)Correlation between $L_{\gamma}$ versus $\Gamma_{c}$, (b)
 $L_{\gamma}$ versus $\Gamma_{c}w_{IR}$ and (c) $L_{\gamma}$ versus $\Gamma_{c}w_{opt}$. Data obtained from Abdo et al. 2010b. }
\label{correlation}
\end{figure}

\acknowledgments

KSC is supported by a GRF grant of Hong Kong Government under HKU700908P, DOC and VAD  are partly supported by the RFBR grant 08-02-00170-a,
the NSC-RFBR Joint Research Project RP09N04 and
09-02-92000-HHC-a and by the grant of a President of the Russian
Federation "Scientific School of Academician V.L.Ginzburg", and AKHK is supported partly by the National
Science Council of the Republic of China (Taiwan)
through grant NSC96-2112-M007-037-MY3.

\end{document}